# A simple interpolation-based approach towards the development of an accurate phenomenological constitutive relation for isotropic hyperelastic materials

Shun Meng[a], Haroon Imtiaz[a], Bin Liu[a,*]

[a] AML, CNMM, Department of Engineering Mechanics, Tsinghua University, Beijing 100084, China

[*] Corresponding author.

Email addresses: liubin@tsinghua.edu.cn (B. Liu).

**Abstract**

Soft materials such as rubber and hydrogels are commonly used in industry for their excellent hyperelastic behaviour. There are various types of constitutive models for soft materials, and phenomenological models are very popular for finite element method (FEM) simulations. However, it is not easy to construct a model that can precisely predict the complex behaviours of soft materials. In this paper, we suggest that the strain energy functions should be expressed as functions of ordered principal stretches, which have more flexible expressions and are capable of matching various experimental curves. Moreover, the feasible region is small, and simple experiments, such as uniaxial tension/compression and hydrostatic tests, are on its boundaries. Therefore, strain energy functions can be easily constructed by the interpolation of experimental curves, which does not need initial guessing in the form of the strain energy function as most existing phenomenological models do. The proposed strain energy functions are perfectly consistent with the available experimental curves for interpolation. It is found that for incompressible materials, the function via an interpolation from two experimental curves can already predict other experimental curves reasonably well.  To further improve the accuracy, additional experiments can be used in the interpolation.

## 1   Introduction

Soft materials have long been used in industry and daily life due to their large deformation capacity. In recent years, these materials have become increasingly popular in the latest technologies, such as soft robots (Bartlett et al., 2015; Martinez, Glavan, Keplinger, Oyetibo, & Whitesides, 2014; Shepherd et al., 2011) and soft electronics (Kim, Ghaffari, Lu, & Rogers, 2012; Muth et al., 2014; Rogers, Someya, & Huang, 2010). With the wide application of soft materials, it is essential to develop constitutive models that can accurately predict the deformation behaviours. Researchers have developed various constitutive models for soft materials, which can be classified as statistical models and phenomenological models.

The statistical models include the Gaussian-chain model (Wall, 1942), three-chain model (H. M. James & Guth, 1943), four-chain model (Flory, 1950; Flory & Rehner Jr, 1943), and eight-chain models (Arruda & Boyce, 1993; Kroon, 2011) and some models in other forms (Xiang et al., 2018). It is important to note that some statistical models are complicated and are not convenient to use in finite element method (FEM) simulations.

Based on continuous mechanics, Rivlin (Rivlin, 1948) proposed a phenomenological model where the general form of the strain energy function is expressed as the series expansion of the invariants of the stretch tensor. Here, the first term of the expansion series is known as the neo-Hookean model. Mooney (Mooney, 1940) derived another form of the phenomenological model that is equivalent to the first two terms of Rivlin's model and has a constant shear modulus. To improve the performance of the phenomenological models, researchers have further investigated the Rivlin model within the continuum mechanics framework for the strain energy function and incorporated higher-order terms of the invariants in the model (Haupt & Sedlan, 2001; Isihara, Hashitsume, & Tatibana, 1951; A. James, Green, & Simpson, 1975; Lion, 1997; Yeoh, 1990, 1993). Furthermore, other forms of the invariant-based strain energy function, which are different from Rivlin's model, have been proposed (Bahreman & Darijani, 2015; Carroll, 2011; Fung, 1967; Gent, 1996; Gent & Thomas, 1958; Khajehsaeid, Arghavani, & Naghdabadi, 2013; Mansouri & Darijani, 2014; Yeoh & Fleming, 1997). On the other hand, Valanis and Landel (Valanis & Landel, 1967) suggested that the strain energy function can be expressed as a function of principal stretches. Following Valanis' research, Ogden (Ogden, 1972; Twizell & Ogden, 1983) proposed a form of model that is a polynomial function of three principal stretches. Similarly, other researchers then developed various models that are expressed as a function of the principal stretches (Darijani & Naghdabadi, 2010). In recent years, many researchers have focused on visco-hyperelastic constitutive models (Lu, Wang, Yang, & Wang, 2017; Upadhyay, Subhash, & Spearot, 2020; Xiang et al., 2019) and anisotropic hyperelastic models (Brown & Smith, 2011; O'Shea, Attard, & Kellermann, 2019; Zhong et al., 2019) due to the rapid development of materials science.

It is important to note from most existing constitutive models that one needs to assume the expression of the strain energy function first for the principal-stretch and invariant-based models. The performance of these models largely depends on the initial guess of the form of the strain energy function. Furthermore, the prediction of some of these models is not always consistent with the experimental data that are used in determining the parameters (Boyce & Arruda, 2000).

Considering the shortcomings of existing models, we hereby propose a straightforward approach to construct a more accurate phenomenological constitutive model for isotropic hyperelastic materials without guessing the form of the strain energy function. In section 2, we propose a constitutive model that employs the ordered principal stretches as independent variables for the strain energy function. Furthermore, we construct the strain energy function by interpolation for incompressible and compressible hyperelastic materials in section 3 and section 4, respectively. In section 5, we discuss a method to further improve the accuracy of the proposed model by incorporating more experimental data. Section 6 concludes this work.

## 2 Proposed strain energy function and a comparison with existing constitutive models

In this section, we first discuss two types of existing phenomenological models for isotropic hyperelastic materials. Then, we propose a different type of strain energy function based on the ordered principal stretches to overcome their disadvantages.

### 2.1 Strain energy function based on invariants

For isotropic hyperelastic materials, the strain energy function is required to be an isotropic function in terms of three principal stretches (Valanis & Landel, 1967). This requirement implies that the exchange of variables does not change the expression of the function, which is a mathematical description of the isotropic property.

As the invariants of the stretch tensor are isotropic functions, it is straightforward to express the strain energy function in terms of the invariants.

$$\widehat{W} = \widehat{W}(I_1, I_2, I_3) \tag{1}$$

where $I_1, I_2,$ and $I_3$ are three invariants of the stretch tensor. Clearly, this type of strain energy function is naturally an isotropic function in terms of the three principal stretches. However, it is difficult to express some isotropic functions of the three principal stretches, such as the example given in 2.3.

### 2.2 Strain energy function based on the principal stretches

To achieve a more flexible expression, one may adopt the principal-stretch-based strain energy function.

$$\widetilde{W} = \widetilde{W}(\lambda_1, \lambda_2, \lambda_3) \tag{2}$$

where $\lambda_1, \lambda_2,$ and $\lambda_3$ are three principal stretches, and the subscripts 1, 2, and 3 represent the three principal directions. It is noted that this type of strain energy function should be a symmetrical expression in terms of the principal stretches (Valanis & Landel, 1967) to satisfy the isotropic requirement. Therefore, this requirement poses a limitation and brings inconvenience in constructing the strain energy function.

### 2.3 Strain energy function based on ordered principal stretches

Considering the shortcomings of existing phenomenological models, we propose a form of phenomenological model in terms of ordered principal stretches.

$$W = W(\lambda_{max}, \lambda_{mid}, \lambda_{min}) \tag{3}$$

where $\lambda_{max}, \lambda_{mid},$ and $\lambda_{min}$ are the maximum, middle and minimum principal stretches, respectively.

The main advantages of the strain energy function based on ordered principal stretches are listed as follows:
- Since $\lambda_{max}, \lambda_{mid}$, and $\lambda_{min}$ are isotropic functions of $\lambda_1, \lambda_2,$ and $\lambda_3$, respectively, the strain energy function in this form is naturally an isotropic function.
- This form of the strain energy function provides more flexibility than invariant-based and principal stretch-based strain energy functions. For instance, the strain energy function in the proposed form, $W = (\lambda_{max}\lambda_{mid}\lambda_{min} - 1)^2 + \lambda_{max}(\lambda_{max} - \lambda_{mid})^2(\lambda_{mid} - \lambda_{min})^2$ is very difficult to express by Eqs. (1)-(2). By contrast, an invariant-based or principal-stretch-based strain energy function can be easily converted into the proposed form.
- Note that $\lambda_{max} \geq \lambda_{mid} \geq \lambda_{min}$, so that the feasible region is smaller than that of the principal-stretches-based strain energy function, as shown in Fig. 2 and Fig. 4, which makes it relatively easier to construct a strain energy function.

The smaller feasible region in the space of the ordered principal stretches is obtained by taking advantage of the exchange symmetry of $\lambda_1, \lambda_2, \lambda_3$. When $\lambda_1 \geq \lambda_2 \geq \lambda_3$, $W(\lambda_{max}, \lambda_{mid}, \lambda_{min})$ equals $\widetilde{W}(\lambda_1, \lambda_2, \lambda_3)$. As $\widetilde{W}(\lambda_1, \lambda_2, \lambda_3)$ is smooth in the $\lambda_1 - \lambda_2 - \lambda_3$ space, continuity is required at the boundary planes $\lambda_{max} = \lambda_{mid}$ and $\lambda_{mid} = \lambda_{min}$:

$$\left.\frac{\partial \widetilde{W}}{\partial \lambda_1}\right|_{\lambda_1 = \lambda_2^+} = \left.\frac{\partial \widetilde{W}}{\partial \lambda_1}\right|_{\lambda_1 = \lambda_2^-}, \quad \left.\frac{\partial \widetilde{W}}{\partial \lambda_2}\right|_{\lambda_2 = \lambda_3^+} = \left.\frac{\partial \widetilde{W}}{\partial \lambda_2}\right|_{\lambda_2 = \lambda_3^-} \tag{4}$$

where $\left.\frac{\partial \widetilde{W}}{\partial \lambda_1}\right|_{\lambda_1 = \lambda_2^+}$ represents the limit of $\frac{\partial \widetilde{W}}{\partial \lambda_1}$ as $\lambda_1$ decreases in value approaching $\lambda_2$ and $\left.\frac{\partial \widetilde{W}}{\partial \lambda_1}\right|_{\lambda_1 = \lambda_2^-}$ represents the limit of $\frac{\partial \widetilde{W}}{\partial \lambda_1}$ as $\lambda_1$ increases in value approaching $\lambda_2$.

Therefore, we have the following continuity conditions for $W(\lambda_{max}, \lambda_{mid}, \lambda_{min})$

$$\left.\frac{\partial W}{\partial \lambda_{max}}\right|_{\lambda_{max} = \lambda_{mid}} = \left.\frac{\partial W}{\partial \lambda_{mid}}\right|_{\lambda_{max} = \lambda_{mid}}, \quad \left.\frac{\partial W}{\partial \lambda_{mid}}\right|_{\lambda_{mid} = \lambda_{min}} = \left.\frac{\partial W}{\partial \lambda_{min}}\right|_{\lambda_{mid} = \lambda_{min}} \tag{5}$$

These continuity conditions can also be interpreted by considering the physical meaning. Along the principal axes, the partial derivatives of the strain energy function with respect to principal stretches are the nominal stresses $P_i$; therefore, Eq. (5) becomes

$$P_{max}|_{\lambda_{max} = \lambda_{mid}} = \left.\frac{\partial W}{\partial \lambda_{max}}\right|_{\lambda_{max} = \lambda_{mid}} = \left.\frac{\partial W}{\partial \lambda_{mid}}\right|_{\lambda_{max} = \lambda_{mid}} = P_{mid}|_{\lambda_{max} = \lambda_{mid}}$$

$$P_{mid}|_{\lambda_{mid} = \lambda_{min}} = \left.\frac{\partial W}{\partial \lambda_{mid}}\right|_{\lambda_{mid} = \lambda_{min}} = \left.\frac{\partial W}{\partial \lambda_{min}}\right|_{\lambda_{mid} = \lambda_{min}} = P_{min}|_{\lambda_{mid} = \lambda_{min}} \tag{6}$$

which just reflects the isotropic property when $\lambda_{max} = \lambda_{mid}$ and $\lambda_{mid} = \lambda_{min}$. The definition of the nominal stress is shown in Fig. 1. It should be pointed out that all nominal stresses and principal stretches can be directly obtained from experiments.

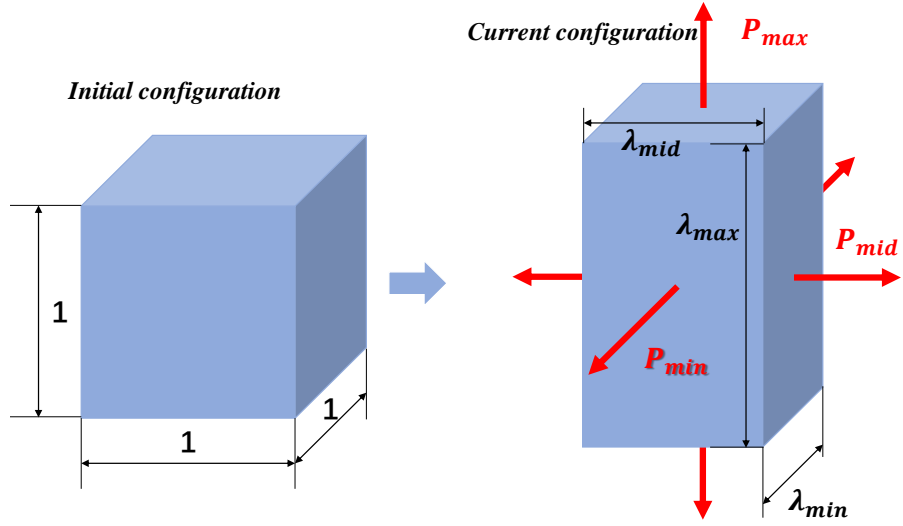

Fig. 1 Schematic diagram for the nominal stress

With this flexible form of the strain energy function, it is possible to construct a phenomenological constitutive relation by the interpolation of available experimental stress-strain curves. In the next two sections, we will demonstrate that for incompressible and compressible isotropic hyperelastic materials, an approximate strain energy functions in terms of the ordered principal stretches can be obtained directly from several simple tests, such as uniaxial tension, uniaxial compression and hydrostatic tests. Specifically, this straightforward approach does not need to guess the form of the function.

## 3  Construction of an incompressible isotropic hyperelastic constitutive model
### 3.1  Feasible region

There are only two independent variables for incompressible materials considering the incompressible restriction

$$\lambda_{max} \lambda_{mid} \lambda_{min} = 1 \tag{7}$$

We denote $W_{incom}(\lambda_{max}, \lambda_{mid})$ as the strain energy function for incompressible cases. Considering the relation $\lambda_{max} \geq \lambda_{mid} \geq \lambda_{min}$ and Eq. (7), we can obtain

$$\lambda_{mid} \geq \lambda_{min} = \frac{1}{\lambda_{mid} \lambda_{max}} \tag{8}$$

which can be rewritten as

$$\lambda_{mid} \geq \lambda_{max}^{-\frac{1}{2}} \tag{9}$$

As shown in Fig. 2, the feasible region for $W_{incom}(\lambda_{max}, \lambda_{mid})$ in $\lambda_{max} - \lambda_{mid}$ space is thus

$$\lambda_{max} \geq \lambda_{mid} \geq \lambda_{max}^{-\frac{1}{2}} \tag{10}$$

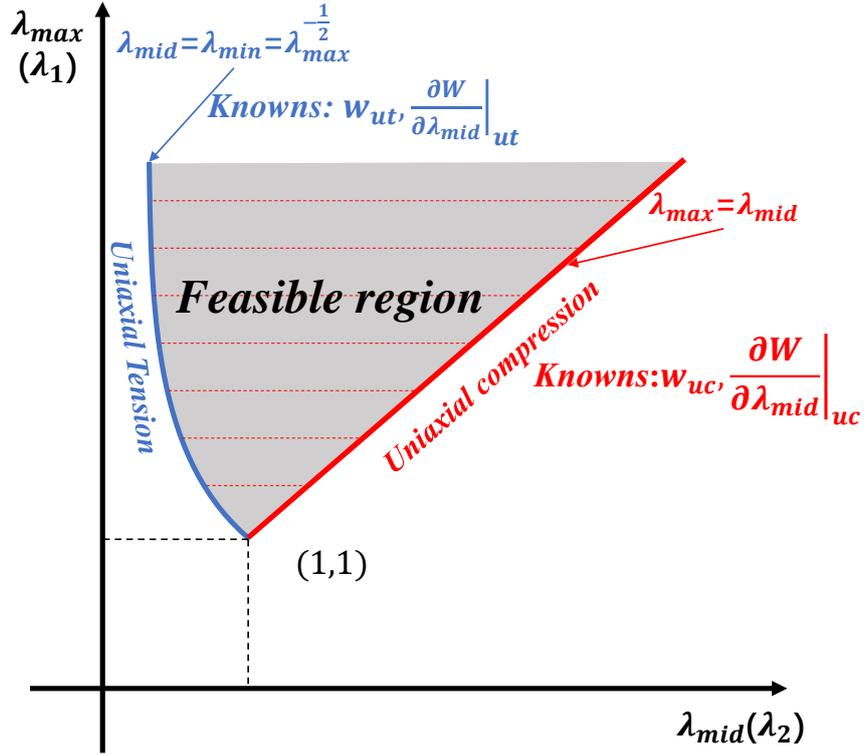

Fig. 2 Feasible region for the incompressible strain energy function

The left boundary (i.e., $\lambda_{mid} = \lambda_{min} = \lambda_{max}^{-\frac{1}{2}}$) corresponds to uniaxial tension or equibiaxial compression, so it is denoted as a uniaxial tension boundary hereafter. Similarly, the right boundary (i.e., $\lambda_{max} = \lambda_{mid}$) corresponds to uniaxial compression or equibiaxial tension and is denoted as the uniaxial compression boundary hereafter.

### 3.2 Boundary conditions for constructing a strain energy function

The strain energy at the uniaxial tension boundary can be defined as

$$w_{ut}(\lambda_{max}) \equiv W_{incom}(\lambda_{max}, \lambda_{mid})\big|_{\lambda_{mid}=\lambda_{min}=\lambda_{max}^{-\frac{1}{2}}} \tag{11}$$

which can be obtained from the uniaxial tension

$$w_{ut}(\lambda_{max}) = \int_1^{\lambda_{max}} P_{max} d\lambda_{max} \tag{12}$$

The subscript $ut$ in this paper represents the uniaxial tension. Note that only one independent variable $\lambda_{max}$ is needed to characterize the deformation of a uniaxial tension for incompressible isotropic hyperelastic materials.

Similarly, the strain energy at the uniaxial compression boundary can be defined as

$$w_{uc}(\lambda_{max}) \equiv W_{incom}(\lambda_{max}, \lambda_{mid})\big|_{\lambda_{mid}=\lambda_{max}} \tag{13}$$

and can be obtained from uniaxial compression

$$w_{uc}(\lambda_{max}) = \tilde{w}_{uc}(\lambda_{min}) = \int_1^{\lambda_{min}} P_{min} d\lambda_{min}$$
$$= \int_1^{\lambda_{max}} P_{min} d\lambda_{max}^{-2} = \int_1^{\lambda_{max}} -2\lambda_{max}^{-3} P_{min} d\lambda_{max} \tag{14}$$

In the derivation above, $\lambda_{min} = \lambda_{max}^{-2}$ for the uniaxial compression test is used. The subscript $uc$ in this paper represents the uniaxial compression.

Regarding the derivatives of the strain energy, the following relation holds for the whole feasible region of $W_{incom}$

$$\frac{\partial W_{incom}}{\partial \lambda_{mid}} = P_{mid} - \frac{1}{\lambda_{max} \lambda_{mid}^2} P_{min} \tag{15}$$

Its derivation is given in Appendix A. Therefore, the partial derivative of $W_{incom}$ with respect to $\lambda_{mid}$ for experimental data can be obtained directly. For uniaxial tension, $P_{mid} = P_{min} = 0$, so that Eq. (15) becomes

$$\left.\frac{\partial W_{incom}}{\partial \lambda_{mid}}\right|_{ut} = \left.\frac{\partial W_{incom}}{\partial \lambda_{mid}}\right|_{\lambda_{mid} = \lambda_{min} = \lambda_{max}^{-\frac{1}{2}}} = 0 \tag{16}$$

At the uniaxial compression boundary, $\lambda_{max} = \lambda_{mid} = \lambda_{min}^{-1/2}$, and for uniaxial compression $P_{mid} = 0$, so that Eq. (15) becomes

$$\left.\frac{\partial W_{incom}}{\partial \lambda_{mid}}\right|_{uc} = \left.\frac{\partial W_{incom}}{\partial \lambda_{mid}}\right|_{\lambda_{max} = \lambda_{mid}} = -\frac{1}{\lambda_{max} \lambda_{mid}^2} P_{min} = -\lambda_{max}^{-3} P_{min} \tag{17}$$

From the derivation above, the boundary conditions can be obtained directly from the uniaxial tension and compression tests. Actually, the boundary conditions can also be obtained from other experiments, such as equibiaxial compression or equibiaxial tension. The strain energy for each experiment can be obtained by

$$w = \int P_{max} d\lambda_{max} + P_{mid} d\lambda_{mid} + P_{min} d\lambda_{min}$$
$$= \int_1^{\lambda_{max}} \left( P_{max} + P_{mid} \cdot \frac{d\lambda_{mid}}{d\lambda_{max}} + P_{min} \cdot \frac{d\lambda_{min}}{d\lambda_{max}} \right) d\lambda_{max} \tag{18}$$

and the corresponding partial derivative can be obtained from Eq. (15). For more details, see Appendix A.

### 3.3 Interpolation for constructing a strain energy function

As shown in Fig. 2, the strain energy function and the derivative for incompressible materials at the two boundaries are obtained from two simple tests. The strain energy function inside the feasible region can be approximated by interpolation between two boundaries along the $\lambda_{mid}$-direction.

One requirement should be satisfied that the strain energy function obtained by the interpolation can degenerate into the following strain energy function for infinitesimal deformation in terms of the principal stretches:

$$W_{incom} = \frac{2E}{3}\left[(\lambda_{max}-1)^2 + (\lambda_{max}-1)(\lambda_{mid}-1) + (\lambda_{mid}-1)^2\right] \tag{19}$$

where $E$ is Young's modulus. The detailed derivation is given in Appendix B. Eq. (19) is a quadratic polynomial function, so the polynomial function will be used during interpolation in this paper. As there are four known conditions from two boundaries, i.e., $w_{ut}, w_{uc}, \left.\frac{\partial W_{incom}}{\partial \lambda_{mid}}\right|_{ut}, \left.\frac{\partial W_{incom}}{\partial \lambda_{mid}}\right|_{uc}$, the strain energy function can be constructed using cubic polynomial interpolation:

$$W_{incom} = a_0(\lambda_{max}) + a_1(\lambda_{max})\cdot\lambda_{mid} + a_2(\lambda_{max})\cdot\lambda_{mid}^2 + a_3(\lambda_{max})\cdot\lambda_{mid}^3 \tag{20}$$

The partial derivative of Eq. (20) with respect to $\lambda_{mid}$ can be expressed as

$$\frac{\partial W_{incom}}{\partial \lambda_{mid}} = a_1(\lambda_{max}) + 2a_2(\lambda_{max})\cdot\lambda_{mid} + 3a_3(\lambda_{max})\cdot\lambda_{mid}^2 \tag{21}$$

Substituting Eqs. (12),(14),(16), and (17) into Eqs. (20)-(21), the parameters $a_0, a_1, a_2, a_3$ of Eq. (20) can then be determined.

Actually determining Eq. (20) can also be regarded as an Hermite interpolation problem:

$$W_{incom} = \alpha_1 w_{ut} + \beta_1 \left.\frac{\partial W_{incom}}{\partial \lambda_{mid}}\right|_{ut} + \alpha_2 w_{uc} + \beta_2 \left.\frac{\partial W_{incom}}{\partial \lambda_{mid}}\right|_{uc} \tag{22}$$

where

$$\alpha_1 = \left(1 + 2\frac{\lambda_{mid}-\lambda_{mid}^L}{\lambda_{mid}^U-\lambda_{mid}^L}\right)\left(\frac{\lambda_{mid}-\lambda_{mid}^U}{\lambda_{mid}^L-\lambda_{mid}^U}\right)^2, \quad \alpha_2 = \left(1 + 2\frac{\lambda_{mid}-\lambda_{mid}^U}{\lambda_{mid}^L-\lambda_{mid}^U}\right)\left(\frac{\lambda_{mid}-\lambda_{mid}^L}{\lambda_{mid}^U-\lambda_{mid}^L}\right)^2 \tag{23}$$

$$\beta_1 = (\lambda_{mid}-\lambda_{mid}^L)\left(\frac{\lambda_{mid}-\lambda_{mid}^U}{\lambda_{mid}^L-\lambda_{mid}^U}\right)^2, \quad \beta_2 = (\lambda_{mid}-\lambda_{mid}^U)\left(\frac{\lambda_{mid}-\lambda_{mid}^L}{\lambda_{mid}^U-\lambda_{mid}^L}\right)^2 \tag{24}$$

where $\lambda_{mid}^L = \lambda_{max}^{-\frac{1}{2}}, \lambda_{mid}^U = \lambda_{max}$ in Eqs. (23)-(24).

### 3.4 Validation of the proposed strain energy function

Now, we validate the strain energy function with the experimental data from Treloar (Treloar, 1944). Following the proposed construction approach, especially Eq. (15) and Eqs. (18) and (22), an approximate strain energy function can be obtained by interpolation between the uniaxial tension and equibiaxial tension curves shown in Fig. 3(a). The predicted pure shear curve from the strain energy function is shown in Fig. 3(b), which agrees well with the experimental data by Treloar, and the correlation coefficient is 0.9186. It should be emphasized that no fitting parameter or arbitrary guess is adopted in the construction of the energy function.

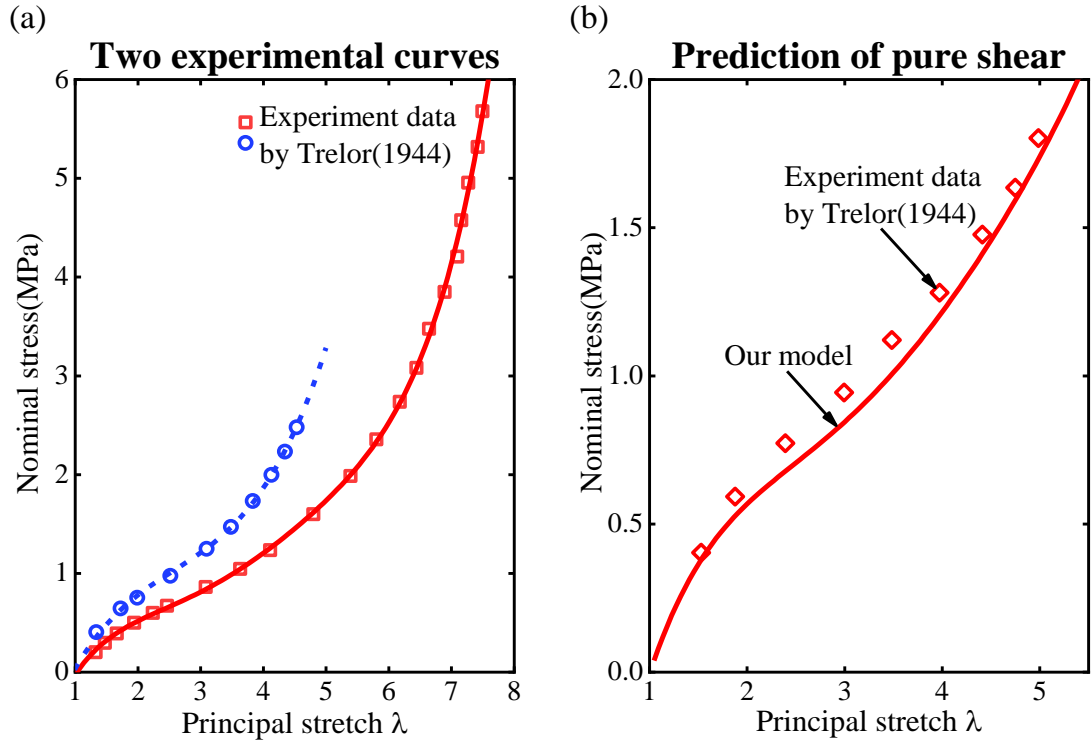

Fig. 3 (a) Two simple experiments used for the construction of the strain energy function; (b) Prediction of the pure shear with the proposed strain energy function

## 4 Construction of the compressible hyperelastic constitutive model

We have discussed the construction of the strain energy function for incompressible hyperelastic materials in the previous section. Compressible hyperelastic materials such as hydrogels are also a class of important materials. In this section, we will discuss the construction of a strain energy function for compressible hyperelastic materials.

### 4.1 Feasible region

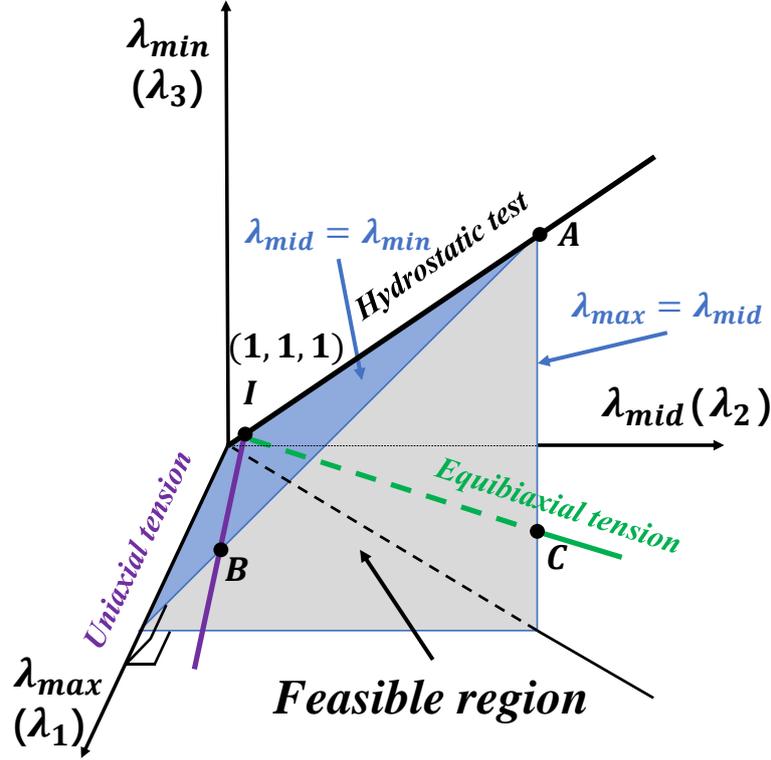

Fig. 4 Feasible region for a compressible strain energy function

The feasible region for compressible materials can be obtained according to $\lambda_{max} \geq \lambda_{mid} \geq \lambda_{min} > 0$, as shown in Fig. 4. The boundaries are composed of three planes: $\lambda_{max} = \lambda_{mid}$, $\lambda_{mid} = \lambda_{min}$, and $\lambda_{min} = 0$. Similar to the incompressible cases, the strain energy for compressible cases will be constructed by interpolation using the boundary conditions. However, for compressible materials, a finite number of experimental curves cannot cover the boundary planes. Therefore, we need to first construct the strain energy function on the boundary planes $\lambda_{mid} = \lambda_{min}$ (the strain energy of this boundary is denoted as $\hat{w}_{AB}$) and $\lambda_{max} = \lambda_{mid}$ (the strain energy of this boundary is denoted as $\hat{w}_{AC}$). At least three experimental curves are required to perform interpolation on the two boundary planes.

Three simple experiments are suggested here: A. Hydrostatic test. B. Uniaxial tension test (or equibiaxial compression test). C. Equibiaxial tension test (or uniaxial compression test). The first experiment is the line of the intersection of plane $\lambda_{max} = \lambda_{mid}$ and plane $\lambda_{mid} = \lambda_{min}$. The second experiment lies on plane $\lambda_{mid} = \lambda_{min}$, and the third experiment is on plane $\lambda_{max} = \lambda_{mid}$. The intersection point (denoted as $I$ in Fig. 4) of the three experimental lines represents the initial configuration. The strain energy functions for the three experiments $w_{hs}$ (corresponding to the hydrostatic test), $w_{ut}$ and $w_{bt}$ (corresponding to the equibiaxial tension test) can be obtained from Eq. (18), and the partial derivatives are the corresponding nominal stresses:

$$\left.\frac{\partial W}{\partial \lambda_{max}}\right|_i = P_{max}^{(i)}, \left.\frac{\partial W}{\partial \lambda_{mid}}\right|_i = P_{mid}^{(i)}, \left.\frac{\partial W}{\partial \lambda_{min}}\right|_i = P_{min}^{(i)} \qquad (25)$$

where $i = hs, ut, bt$ for the three simple experiments.

### 4.2 Interpolation for constructing a strain energy function

The interpolation strategy for compressible materials is given as follows: First, we obtain the strain energy and the partial derivatives on the experimental curves. Then, interpolation is performed to obtain the strain energy functions on the boundary planes, i.e. $\hat{w}_{AB}(\lambda_{max}, \lambda_{min})$ and $\hat{w}_{AC}(\lambda_{max}, \lambda_{min})$. Finally, the strain energy function in the whole feasible region $W(\lambda_{max}, \lambda_{mid}, \lambda_{min})$ can be approximated by interpolation along the $\lambda_{mid}$-direction between these two boundary planes. The interpolation strategy is schematically shown in Fig. 5, and the details are given in the following.

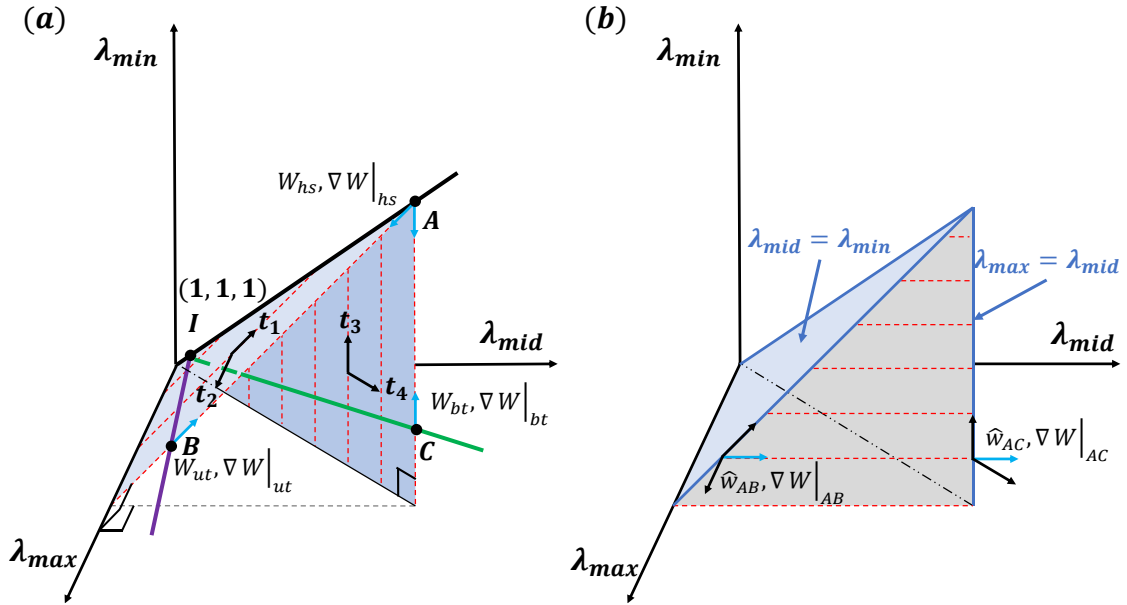

Fig. 5 Interpolation approach for the compressible strain energy function: (a) interpolation to obtain boundaries and (b) interpolation to obtain the strain energy function

***Step 1: Interpolation from experimental curves to boundary planes***

For a specific $\lambda_{max}$, $\hat{w}_{AB}$ and $\hat{w}_{AC}$ can be approximated by interpolation with respect to $\lambda_{min}$ along the red dashed line shown in Fig. 5(a). To obtain the related derivatives, the variation of the strain energy function can be written as:

$$\delta W = P_{max} \cdot \delta\lambda_{max} + P_{mid} \cdot \delta\lambda_{mid} + P_{min} \cdot \delta\lambda_{min} \qquad (26)$$

For plane $AIB$, $\lambda_{mid} = \lambda_{min}$, $P_{mid} = P_{min}$, and $W$ degenerates into $\hat{w}_{AB}$; therefore, the above equation becomes

$$\delta\hat{w}_{AB} = P_{max}\big|_{\lambda_{mid}=\lambda_{min}} \cdot \delta\lambda_{max} + 2P_{min}\big|_{\lambda_{mid}=\lambda_{min}} \cdot \delta\lambda_{min} \qquad (27)$$

$$\frac{\partial \hat{w}_{AB}}{\partial \lambda_{min}} = 2P_{min} \qquad (28)$$

For uniaxial tension and hydrostatic experiment curves, $\hat{w}_{AB}$ and its derivatives are known, and for the whole plane $AIB$, $\hat{w}_{AB}$ can be constructed by interpolation:

$$\begin{aligned}\hat{w}_{AB}(\lambda_{max},\lambda_{min}) &= \alpha_1^{AB} w_{ut} + \beta_1^{AB} \frac{\partial \hat{w}_{AB}}{\partial \lambda_{min}}\bigg|_{ut} + \alpha_2^{AB} w_{hs} + \beta_2^{AB} \frac{\partial \hat{w}_{AB}}{\partial \lambda_{min}}\bigg|_{hs} \\ &= \alpha_1^{AB} w_{ut} + 2\beta_1^{AB} P_{min}^{(ut)} + \alpha_2^{AB} w_{hs} + 2\beta_2^{AB} P_{min}^{(hs)}\end{aligned} \qquad (29)$$

where $\alpha_1^{AB}, \beta_1^{AB}, \alpha_2^{AB}, \beta_2^{AB}$ are given in Eq.(D-8).

For $\hat{w}_{AC}$ on plane $AIC$, substituting $\lambda_{max} = \lambda_{mid}$ into Eq. (26) yields

$$\frac{\partial \hat{w}_{AC}}{\partial \lambda_{min}} = P_{min} \qquad (30)$$

$\hat{w}_{AC}$ can then be constructed by interpolation:

$$\begin{aligned}\hat{w}_{AC}(\lambda_{max},\lambda_{min}) &= \alpha_1^{AC} w_{bt} + \beta_1^{AC} \frac{\partial \hat{w}_{AC}}{\partial \lambda_{min}}\bigg|_{bt} + \alpha_2^{AC} w_{hs} + \beta_2^{AC} \frac{\partial \hat{w}_{AC}}{\partial \lambda_{min}}\bigg|_{hs} \\ &= \alpha_1^{AC} w_{bt} + \beta_1^{AC} P_{min}^{(bt)} + \alpha_2^{AC} w_{hs} + \beta_2^{AC} P_{min}^{(hs)}\end{aligned} \qquad (31)$$

where $\alpha_1^{AC}, \beta_1^{AC}, \alpha_2^{AC}, \beta_2^{AC}$ are given in Eq.(D-9).

***Step 2: Interpolation from the boundary planes to the whole feasible region***

Interpolation along the $\lambda_{mid}$-direction (the red dashed line shown in Fig. 5(b)) between two boundary planes is implemented to obtain the strain energy function over a whole feasible region. For the boundary plane $\lambda_{mid} = \lambda_{min}$, the partial derivative $\left.\frac{\partial W}{\partial \lambda_{mid}}\right|_{\lambda_{mid}=\lambda_{min}}$ can be obtained by considering Eq. (28):

$$\left.\frac{\partial W}{\partial \lambda_{mid}}\right|_{\lambda_{mid}=\lambda_{min}} = \left. P_{mid}\right|_{\lambda_{mid}=\lambda_{min}} = \left. P_{min}\right|_{\lambda_{mid}=\lambda_{min}} = \frac{1}{2}\frac{\partial \hat{w}_{AB}}{\partial \lambda_{min}} \tag{32}$$

where $\frac{\partial \hat{w}_{AB}}{\partial \lambda_{min}}$ can be obtained by the partial derivation of Eq. (29). Similarly, $\left.\frac{\partial W}{\partial \lambda_{mid}}\right|_{\lambda_{mid}=\lambda_{max}}$ can be obtained by considering Eq. (26)

$$\delta \hat{w}_{AC} = 2\left. P_{max}\right|_{\lambda_{mid}=\lambda_{max}} \cdot \delta \lambda_{max} + \left. P_{min}\right|_{\lambda_{mid}=\lambda_{max}} \cdot \delta \lambda_{min} \tag{33}$$

$$\frac{\partial \hat{w}_{AC}}{\partial \lambda_{max}} = 2 P_{max} \tag{34}$$

$$\left.\frac{\partial W}{\partial \lambda_{mid}}\right|_{\lambda_{mid}=\lambda_{max}} = \left. P_{mid}\right|_{\lambda_{mid}=\lambda_{max}} = \left. P_{max}\right|_{\lambda_{mid}=\lambda_{max}} = \frac{1}{2}\frac{\partial \hat{w}_{AC}}{\partial \lambda_{max}} \tag{35}$$

Then, the strain energy function for compressible cases can be approximated by constructing an interpolation function with respect to $\lambda_{mid}$.

$$\begin{aligned} W(\lambda_{max}, \lambda_{mid}, \lambda_{min}) &= \alpha_1^{com} \left. W\right|_{\lambda_{mid}=\lambda_{min}} + \beta_1^{com} \left.\frac{\partial W}{\partial \lambda_{mid}}\right|_{\lambda_{mid}=\lambda_{min}} + \alpha_2^{com} \left. W\right|_{\lambda_{mid}=\lambda_{max}} + \beta_2^{com} \left.\frac{\partial W}{\partial \lambda_{mid}}\right|_{\lambda_{mid}=\lambda_{max}} \\ &= \alpha_1^{com} \hat{w}_{AB} + \frac{\beta_1^{com}}{2}\frac{\partial \hat{w}_{AB}}{\partial \lambda_{min}} + \alpha_2^{com} \hat{w}_{AC} + \frac{\beta_2^{com}}{2}\frac{\partial \hat{w}_{AC}}{\partial \lambda_{max}} \end{aligned}$$

(36)

where $\alpha_1^{com}, \beta_1^{com}, \alpha_2^{com}, \beta_2^{com}$ are given in Eq.(D-11).

It should be noted that the obtained strain energy function satisfies the continuity conditions automatically because the conditions are used in the derivation of Eqs. (32)-(35).

## 5  More experiments to improve accuracy

In this work, we use two simple experiments to construct the strain energy function for incompressible materials, whereas at least three simple experiments are required for compressible materials. However, this is the basic requirement for constructing the strain energy function. More experiments might be needed for more accurate strain energy functions. In the following, we discuss how to construct a strain energy function with additional experiments.

We take incompressible materials as an example. $P_{max}, P_{mid}, P_{min}$, $\lambda_{mid}, \lambda_{min}$ and $\lambda_{max}$ can be measured and recorded for a certain type of experiment denoted as $i$. Its strain energy $w_i$ can be obtained using Eq. (18) and the partial derivative can also be derived from Eq. (15). To make use of these additional conditions, as shown in Fig. 6, the strain energy function can be approximated by higher-order interpolation or piecewise interpolation in the divided regions.

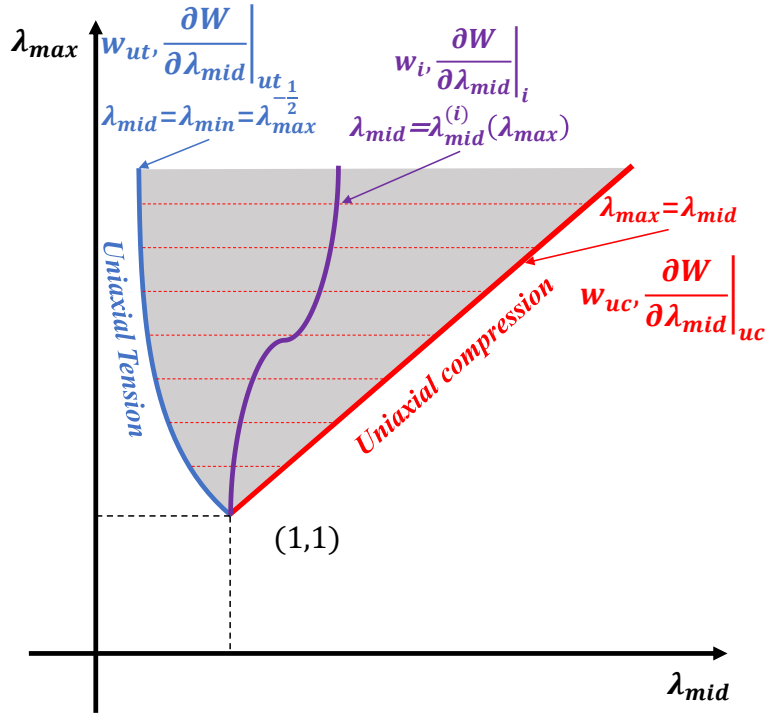

Fig. 6 Interpolation strategy for incompressible materials for a higher accuracy

## 6  Procedure of constructing the strain energy function

Therefore, the construction procedure for a strain energy function can be summarized into the following steps:

**Step 1:** Obtain $P - \lambda$ curves from the experiments.
**Step 2:** Compute the strain energy function and partial derivatives for each experimental curve
**Step 3:** Interpolation is performed to obtain the strain energy function.
- For incompressible materials, interpolation is performed to obtain the strain energy function from the boundary experimental curves.

- For compressible materials, interpolation is performed from the boundary experimental curves to the boundary planes and then to the whole feasible region.

If the accuracy is not satisfied:

**Step 4:** Additional experiments are performed inside the feasible region and higher-order or piecewise interpolation is carried out to update the strain energy function.

The detailed procedure including equations for construction of strain energy function is given in Appendix C and Appendix D.

## 7  Conclusions

In this work, we propose an approach to construct strain energy functions for isotropic hyperelastic materials. The advantages of this approach over the existing phenomenological models are summarized as follows:

1. We suggest that the strain energy functions should be expressed as functions of the ordered principal stretches, which have more flexible expressions and are capable of matching various experimental curves.
2. The feasible region in the space of the ordered principal stretches is smaller than that in the traditional space of principal stretches, and simple experiments, such as uniaxial tension/compression and hydrostatic tests, are on its boundaries. Therefore, the strain energy functions can be easily constructed by interpolation of these simple experiments.
3. Different from most traditional phenomenological constitutive models, the proposed strain energy functions are perfectly consistent with the available experimental curves for interpolation, and the form of the strain energy functions does not need to be guessed.
4. Our proposed approach for the construction of strain energy functions is validated by experiments from other researchers. The function obtained by interpolation of two experimental curves has already reached an acceptable accuracy.
5. The accuracy of the proposed strain energy functions can be further improved by using more experiments.

**Appendix A: Boundary conditions for the incompressible cases**

During the quasi-static loading process, the strain energy is equal to the work of external forces, so that

$$w = \int P_{max} d\lambda_{max} + P_{mid} d\lambda_{mid} + P_{min} d\lambda_{min}$$
$$= \int_1^{\lambda_{max}} \left( P_{max} + P_{mid} \cdot \frac{d\lambda_{mid}}{d\lambda_{max}} + P_{min} \cdot \frac{d\lambda_{min}}{d\lambda_{max}} \right) d\lambda_{max} \quad \text{(A-1)}$$

$P_{max} - \lambda_{max}, P_{mid} - \lambda_{mid}, P_{min} - \lambda_{min}$ curves can be obtained from the corresponding experiment.

$w_{ut}(\lambda_{max})$ can be obtained from either the uniaxial tension test or equibiaxial compression test or other kinds of experiments that satisfy the restriction $\lambda_{mid} = \lambda_{min} = \lambda_{max}^{-\frac{1}{2}}$ for the uniaxial tension boundary. Then, $w_{ut}(\lambda_{max})$ can be obtained by integration, following Eq. (A-1):

$$w_{ut}(\lambda_{max}) = \int_1^{\lambda_{max}} \left[ P_{max} - \frac{1}{2} \lambda_{max}^{-\frac{3}{2}} \cdot (P_{mid} + P_{min}) \right] d\lambda_{max} \quad \text{(A-2)}$$

Similarly, $w_{uc}(\lambda_{max})$ can be obtained from either the uniaxial compression test or the equibiaxial tension test or other kinds of experiments that satisfy the restriction $\lambda_{min} = \lambda_{mid}^{-2} = \lambda_{max}^{-2}$ for the uniaxial compression boundary:

$$w_{uc}(\lambda_{max}) = \int_1^{\lambda_{max}} \left( P_{max} + P_{mid} - 2\lambda_{max}^{-3} \cdot P_{min} \right) d\lambda_{max} \quad \text{(A-3)}$$

Moreover, the partial derivatives of $W_{incom}$ to $\lambda_{mid}$ at two boundaries can be derived from the continuity conditions. We denote the partial derivative at the uniaxial tension boundary as $\left.\frac{\partial W_{incom}}{\partial \lambda_{mid}}\right|_{ut}$ and the corresponding partial derivatives at the uniaxial compression boundary as $\left.\frac{\partial W_{incom}}{\partial \lambda_{mid}}\right|_{uc}$.

For $W_{incom}(\lambda_{max}, \lambda_{mid})$, using the chain rule

$$\delta W_{incom} = P_{max} \delta \lambda_{max} + P_{mid} \delta \lambda_{mid} + P_{min} \delta \lambda_{min}$$
$$= P_{max} \delta \lambda_{max} + P_{mid} \delta \lambda_{mid} + P_{min} \delta \left( \frac{1}{\lambda_{max} \lambda_{mid}} \right) \quad \text{(A-4)}$$
$$= \left( P_{max} - \frac{1}{\lambda_{max}^2 \lambda_{mid}} P_{min} \right) \delta \lambda_{max} + \left( P_{mid} - \frac{1}{\lambda_{max} \lambda_{mid}^2} P_{min} \right) \delta \lambda_{mid}$$

$$\left.\frac{\partial W_{incom}}{\partial \lambda_{max}}\right|_{\lambda_{mid}} = P_{max} - \frac{1}{\lambda_{max}^2 \lambda_{mid}} P_{min} \quad \text{(A-5)}$$

$$\left.\frac{\partial W_{incom}}{\partial \lambda_{mid}}\right|_{\lambda_{max}} = P_{mid} - \frac{1}{\lambda_{max} \lambda_{mid}^2} P_{min} \quad \text{(A-6)}$$

Then, $\left.\frac{\partial W_{incom}}{\partial \lambda_{mid}}\right|_{ut}$ and $\left.\frac{\partial W_{incom}}{\partial \lambda_{mid}}\right|_{uc}$ can be obtained from Eq. (A-6).

**Appendix B: Strain energy function of infinitesimal deformation theory**

For linear elastic materials undergoing infinitesimal deformation, the strain energy function can be expressed as the sum of the dilatational strain energy and distortional strain energy:

$$W = W_{dila} + W_{dist} = \frac{1}{2} K \left( \varepsilon_{kk} \right)^2 + \frac{E}{2(1+\upsilon)} e_{ij} e_{ij} \tag{B-1}$$

For incompressible materials, $W_{dila} = 0$, $\nu = 0.5$, so we have

$$W_{incom} = W_{dist} = \frac{E}{2(1+\upsilon)} e_{ij} e_{ij} = \frac{E}{3} e_{ij} e_{ij} \tag{B-2}$$

where $E$ is Young's modulus, $\nu$ is Poisson's ratio, and $e_{ij}$ is the deviatoric strain tensor. Along the principal axes:

$$\varepsilon_{ij} = 0, \ i \neq j \tag{B-3}$$

Therefore, the strain energy for incompressible materials becomes

$$W_{incom} = \frac{E}{3} \left( e_{11}^2 + e_{22}^2 + e_{33}^2 \right) = \frac{E}{9} \left[ \left( \varepsilon_{11} - \varepsilon_{22} \right)^2 + \left( \varepsilon_{22} - \varepsilon_{33} \right)^2 + \left( \varepsilon_{33} - \varepsilon_{11} \right)^2 \right] \tag{B-4}$$

Moreover, considering the incompressible requirement that $\varepsilon_{11} + \varepsilon_{22} + \varepsilon_{33} = 0$, the strain energy for incompressible materials becomes

$$W_{incom} = \frac{2E}{3} \left( \varepsilon_{11}^2 + \varepsilon_{11} \varepsilon_{22} + \varepsilon_{22}^2 \right) \tag{B-5}$$

The results above are obtained from infinitesimal deformation theory. Under principal axes, the relation between principal stretches and nominal strain are

$$\varepsilon_{ii} = \lambda_i - 1, \ i = 1, 2, 3 \tag{B-6}$$

so that

$$W_{incom} = \frac{2E}{3} \left[ \left( \lambda_{max} - 1 \right)^2 + \left( \lambda_{max} - 1 \right)\left( \lambda_{mid} - 1 \right) + \left( \lambda_{mid} - 1 \right)^2 \right] \tag{B-7}$$

# Appendix C: Strain energy functions for isotropic hyperelastic materials: Incompressible cases

## Step 1. Obtaining the $P-\lambda$ curves of boundaries

The stress-strain curves during the loading process can be obtained directly from experiment:

$$P_{max}-\lambda_{max}, P_{mid}-\lambda_{mid}, P_{min}-\lambda_{min} \tag{C-1}$$

where $P_{max}, P_{mid}, P_{min}$ are nominal stresses and $\lambda_{max}, \lambda_{mid}, \lambda_{min}$ are the corresponding principal stretches.

## Step 2. Obtaining the strain energy and partial derivatives

The strain energy at the uniaxial tension boundary $w_{ut}(\lambda_{max})$ can be obtained from uniaxial tension test:

$$w_{ut}(\lambda_{max}) = \int_1^{\lambda_{max}} P_{max}^{(ut)}(\lambda_{max})d\lambda_{max} \tag{C-2}$$

or from biaxial compression test:

$$w_{ut}(\lambda_{max}) = -\frac{1}{2}\int_1^{\lambda_{max}} \lambda_{max}^{-\frac{3}{2}} \cdot \left[P_{mid}^{(bc)}(\lambda_{max}) + P_{min}^{(bc)}(\lambda_{max})\right]d\lambda_{max} \tag{C-3}$$

where $P_{max}^{(ut)}(\lambda_{max})$ represents the maximum nominal stress in uniaxial tension, $P_{mid}^{(bc)}(\lambda_{max})$ and $P_{min}^{(bc)}(\lambda_{max})$ represent the middle and minimum nominal stresses in biaxial compression test, which are expressed as the function of $\lambda_{max}$.

The strain energy at the uniaxial compression boundary $w_{uc}$ can be obtained from uniaxial compression test:

$$w_{uc} = -2\int_1^{\lambda_{max}} \lambda_{max}^{-3} \cdot P_{min}^{(uc)}(\lambda_{max})d\lambda_{max} \tag{C-4}$$

or biaxial tension test:

$$w_{uc}(\lambda_{max}) = \int_1^{\lambda_{max}} \left[P_{max}^{(bt)}(\lambda_{max}) + P_{mid}^{(bt)}(\lambda_{max})\right]d\lambda_{max} \tag{C-5}$$

where $P_{min}^{(uc)}(\lambda_{max})$ represents the minimum nominal stress in uniaxial compression test, $P_{max}^{(bt)}(\lambda_{max})$ and $P_{mid}^{(bt)}(\lambda_{max})$ represent the maximum and middle nominal stress in biaxial tension test.

## Step 3. Performing interpolation to obtain strain energy function

For uniaxial compression test or biaxial tension test on the uniaxial compression boundary, the strain energy function can be written as

$$W_{incom} = \begin{cases} \alpha_1 w_{ut} + \alpha_2 w_{uc} - \beta_2 \lambda_{max}^{-3} \cdot P_{min}^{(uc)} & \text{uniaxial tension \& compression} \\ \alpha_1 w_{ut} + \alpha_2 w_{uc} + \beta_2 \cdot P_{mid}^{(bt)} & \text{uniaxial \& biaxial tension} \end{cases} \quad (C-6)$$

where

$$\alpha_1 = \left(1 + 2\frac{\lambda_{mid} - \lambda_{max}^{-1/2}}{\lambda_{max} - \lambda_{max}^{-1/2}}\right)\left(\frac{\lambda_{mid} - \lambda_{max}}{\lambda_{max}^{-1/2} - \lambda_{max}}\right)^2, \quad \alpha_2 = \left(1 + 2\frac{\lambda_{mid} - \lambda_{max}}{\lambda_{max}^{-1/2} - \lambda_{max}}\right)\left(\frac{\lambda_{mid} - \lambda_{max}^{-1/2}}{\lambda_{max} - \lambda_{max}^{-1/2}}\right)^2 \quad (C-7)$$

$$\beta_1 = \left(\lambda_{mid} - \lambda_{max}^{-1/2}\right)\left(\frac{\lambda_{mid} - \lambda_{max}}{\lambda_{max}^{-1/2} - \lambda_{max}}\right)^2, \quad \beta_2 = \left(\lambda_{mid} - \lambda_{max}\right)\left(\frac{\lambda_{mid} - \lambda_{max}^{-1/2}}{\lambda_{max} - \lambda_{max}^{-1/2}}\right)^2 \quad (C-8)$$

With the strain energy function obtained, the nominal stresses can then be derived from Eqs. (A-5)-(A-6) and Eq. (C-6):

$$P_{max} = \begin{cases} \dfrac{P_0}{\lambda_{max}} + \dfrac{\partial \alpha_1}{\partial \lambda_{max}} \int_1^{\lambda_{max}} P_{max}^{(ut)}(\lambda_{max}) d\lambda_{max} - 2\dfrac{\partial \alpha_2}{\partial \lambda_{max}} \int_1^{\lambda_{max}} \dfrac{P_{min}^{(uc)}(\lambda_{max})}{\lambda_{max}^3} d\lambda_{max} \\ + \left(3\lambda_{max}^{-4}\beta_2 - \lambda_{max}^{-3}\left(2\alpha_2 + \dfrac{\partial \beta_2}{\partial \lambda_{max}}\right)\right) P_{min}^{(uc)} - \dfrac{\beta_2}{\lambda_{max}^3} \dfrac{dP_{min}^{(uc)}}{d\lambda_{max}} + \alpha_1 P_{max}^{(ut)} & \begin{array}{l}\text{uniaxial}\\\text{tension}\\\text{\&compression}\end{array} \\ \\ \dfrac{P_0}{\lambda_{max}} + \dfrac{\partial \alpha_1}{\partial \lambda_{max}} \int_1^{\lambda_{max}} P_{max}^{(ut)}(\lambda_{max}) d\lambda_{max} + 2\dfrac{\partial \alpha_2}{\partial \lambda_{max}} \int_1^{\lambda_{max}} P_{max}^{(bt)}(\lambda_{max}) d\lambda_{max} \\ + \left(2\alpha_2 + \dfrac{\partial \beta_2}{\partial \lambda_{max}}\right) P_{max}^{(bt)} + \beta_2 \dfrac{dP_{max}^{(bt)}}{d\lambda_{max}} + \alpha_1 P_{max}^{(ut)} & \begin{array}{l}\text{uniaxial}\\\text{\&biaxial}\\\text{tension}\end{array} \end{cases}$$

(C-9)

$$P_{mid} = \begin{cases} \dfrac{P_0}{\lambda_{mid}} + \dfrac{\partial \alpha_1}{\partial \lambda_{mid}} \int_1^{\lambda_{max}} P_{max}^{(ut)}(\lambda_{max}) d\lambda_{max} \\ -2\dfrac{\partial \alpha_2}{\partial \lambda_{mid}} \int_1^{\lambda_{max}} \dfrac{P_{min}^{(uc)}(\lambda_{max})}{\lambda_{max}^3} d\lambda_{max} - \dfrac{\partial \beta_2}{\partial \lambda_{mid}} \lambda_{max}^{-3} P_{min}^{(uc)} & \begin{array}{l}\text{uniaxial}\\\text{tension}\\\text{\&compression}\end{array} \\ \\ \dfrac{P_0}{\lambda_{mid}} + \dfrac{\partial \alpha_1}{\partial \lambda_{mid}} \int_1^{\lambda_{max}} P_{max}^{(ut)}(\lambda_{max}) d\lambda_{max} \\ +2\dfrac{\partial \alpha_2}{\partial \lambda_{mid}} \int_1^{\lambda_{max}} P_{max}^{(bt)}(\lambda_{max}) d\lambda_{max} + \dfrac{\partial \beta_2}{\partial \lambda_{mid}} P_{max}^{(bt)} & \begin{array}{l}\text{uniaxial}\\\text{\&biaxial}\\\text{tension}\end{array} \end{cases}$$

(C-10)

$$P_{min} = P_0 \lambda_{max} \lambda_{mid} \quad (C-11)$$

where $P_0$ is the minimum Cauchy stress and can be determined by equilibrium condition.

**Step 4. More experiments to improve accuracy**

For a certain kind of experiment denoted as $i$, $P_{max} - \lambda_{max}, P_{mid} - \lambda_{mid}, P_{min} - \lambda_{min}$ curves can be obtained, and the relation between $\lambda_{mid}, \lambda_{min}$ and $\lambda_{max}$ can be expressed as

$$\lambda_{mid} = \lambda_{mid}^{(i)}\left(\lambda_{max}\right), \lambda_{min} = \lambda_{min}^{(i)}\left(\lambda_{max}\right) \tag{C-12}$$

Then, the strain energy $w_i$ can be obtained:

$$w_i = \int_{1}^{\lambda_{max}} \left[ P_{max}^{(i)} + P_{mid}^{(i)} \frac{d\lambda_{mid}^{(i)}}{d\lambda_{max}} + P_{min}^{(i)} \frac{d\lambda_{min}^{(i)}}{d\lambda_{max}} \right] d\lambda_{max} \tag{C-13}$$

The partial derivative can also be derived:

$$\left.\frac{\partial W_{incom}}{\partial \lambda_{mid}}\right|_i = P_{mid}^{(i)} - \frac{1}{\lambda_{max}\left(\lambda_{mid}^{(i)}\right)^2} P_{min}^{(i)} \tag{C-14}$$

Then, higher-order interpolation or piecewise interpolation can be performed to improve the accuracy of the strain energy function.

## Appendix D: Strain energy functions for isotropic hyperelastic materials: Compressible cases

### Step 1. Obtaining $P - \lambda$ curves from the experiments

Similar to the incompressible situations, the stress-strain curves can be obtained:

$$P_{max} - \lambda_{max}, P_{mid} - \lambda_{mid}, P_{min} - \lambda_{min} \tag{D-1}$$

The relation between principal stretches can be expressed as:

$$\lambda_{mid} = \lambda_{mid}^{(i)}(\lambda_{max}), \lambda_{min} = \lambda_{min}^{(i)}(\lambda_{max}) \tag{D-2}$$

The subscript $i$ represents a certain kind of experiment. For hydrostatic test $i = hs$, $i = ut$ and $i = bt$ are similar for uniaxial tension test and equibiaxial tension test.

### Step 2. Obtaining strain energy and partial derivatives

$w_{hs}(\lambda_{max})$ can be obtained from hydrostatic test:

$$w_{hs}(\lambda_{max}) = 3\int_1^{\lambda_{max}} P_{max}^{(hs)}(\lambda_{max}) d\lambda_{max} \tag{D-3}$$

$w_{ut}(\lambda_{max})$ can be obtained from uniaxial tension test:

$$w_{ut}(\lambda_{max}) = \int_1^{\lambda_{max}} P_{max}^{(ut)}(\lambda_{max}) d\lambda_{max} \tag{D-4}$$

$w_{bt}(\lambda_{max})$ can be obtained from uniaxial compression test:

$$w_{bt}(\lambda_{max}) = 2\int_1^{\lambda_{max}} P_{max}^{(bt)}(\lambda_{max}) d\lambda_{max} \tag{D-5}$$

### Step 3. Performing interpolation to construct the strain energy on boundaries

Then $\hat{w}_{AB}$ and $\hat{w}_{AC}$ (shown in Fig. 5) can be constructed by interpolation:

$$\hat{w}_{AB}(\lambda_{max}, \lambda_{min}) = \alpha_1^{AB} w_{ut} + 2\beta_1^{AB} P_{min}^{(ut)} + \alpha_2^{AB} w_{hs} + 2\beta_2^{AB} P_{min}^{(hs)} \tag{D-6}$$

$$\hat{w}_{AC}(\lambda_{max}, \lambda_{min}) = \alpha_1^{AC} w_{bt} + \beta_1^{AC} P_{min}^{(bt)} + \alpha_2^{AC} w_{hs} + \beta_2^{AC} P_{min}^{(hs)} \tag{D-7}$$

where

$$\alpha_1^{AB} = \left(1 + 2\frac{\lambda_{min} - \lambda_{min}^{(ut)}}{\lambda_{max} - \lambda_{min}^{(ut)}}\right)\left(\frac{\lambda_{min} - \lambda_{max}}{\lambda_{min}^{(ut)} - \lambda_{max}}\right)^2, \alpha_2^{AB} = \left(1 + 2\frac{\lambda_{min} - \lambda_{max}}{\lambda_{min}^{(ut)} - \lambda_{max}}\right)\left(\frac{\lambda_{min} - \lambda_{min}^{(ut)}}{\lambda_{max} - \lambda_{min}^{(ut)}}\right)^2$$

$$\beta_1^{AB} = \left(\lambda_{min} - \lambda_{min}^{(ut)}\right)\left(\frac{\lambda_{min} - \lambda_{max}}{\lambda_{min}^{(ut)} - \lambda_{max}}\right)^2, \beta_2^{AB} = \left(\lambda_{min} - \lambda_{max}\right)\left(\frac{\lambda_{min} - \lambda_{min}^{(ut)}}{\lambda_{max} - \lambda_{min}^{(ut)}}\right)^2$$

$$\tag{D-8}$$

$$\alpha_1^{AC} = \left(1 + 2\frac{\lambda_{min} - \lambda_{min}^{(bt)}}{\lambda_{max} - \lambda_{min}^{(bt)}}\right)\left(\frac{\lambda_{min} - \lambda_{max}}{\lambda_{min}^{(bt)} - \lambda_{max}}\right)^2, \quad \alpha_2^{AC} = \left(1 + 2\frac{\lambda_{min} - \lambda_{max}}{\lambda_{min}^{(bt)} - \lambda_{max}}\right)\left(\frac{\lambda_{min} - \lambda_{min}^{(bt)}}{\lambda_{max} - \lambda_{min}^{(bt)}}\right)^2$$

$$\beta_1^{AC} = \left(\lambda_{min} - \lambda_{min}^{(bt)}\right)\left(\frac{\lambda_{min} - \lambda_{max}}{\lambda_{min}^{(bt)} - \lambda_{max}}\right)^2, \quad \beta_2^{AC} = \left(\lambda_{min} - \lambda_{max}\right)\left(\frac{\lambda_{min} - \lambda_{min}^{(bt)}}{\lambda_{max} - \lambda_{min}^{(bt)}}\right)^2$$

(D-9)

### Step 4. Performing interpolation to construct the strain energy function

The strain energy function $W$ can be obtained:

$$W(\lambda_{max}, \lambda_{mid}, \lambda_{min}) = \alpha_1^{com} \hat{w}_{AB} + \frac{\beta_1^{com}}{2}\frac{\partial \hat{w}_{AB}}{\partial \lambda_{min}} + \alpha_2^{com} \hat{w}_{AC} + \frac{\beta_2^{com}}{2}\frac{\partial \hat{w}_{AC}}{\partial \lambda_{max}} \quad \text{(D-10)}$$

where

$$\alpha_1^{com} = \left(1 + 2\frac{\lambda_{mid} - \lambda_{min}}{\lambda_{max} - \lambda_{min}}\right)\left(\frac{\lambda_{mid} - \lambda_{max}}{\lambda_{min} - \lambda_{max}}\right)^2, \quad \alpha_2^{com} = \left(1 + 2\frac{\lambda_{mid} - \lambda_{max}}{\lambda_{min} - \lambda_{max}}\right)\left(\frac{\lambda_{mid} - \lambda_{min}}{\lambda_{max} - \lambda_{min}}\right)^2$$

$$\beta_1^{com} = \left(\lambda_{mid} - \lambda_{min}\right)\left(\frac{\lambda_{mid} - \lambda_{max}}{\lambda_{min} - \lambda_{max}}\right)^2, \quad \beta_2^{com} = \left(\lambda_{mid} - \lambda_{max}\right)\left(\frac{\lambda_{mid} - \lambda_{min}}{\lambda_{max} - \lambda_{min}}\right)^2$$

(D-11)

With the strain energy $W$ obtained, the nominal stresses can then be determined from Eq. (25).

**Acknowledgement**

This work was supported by the National Natural Science Foundation of China [grant numbers 11720101002, 11921002, and 11890674], and Science Challenge Project No. TZ2018001.